\documentclass[twocolumn]{article}

\usepackage[english]{babel}
\usepackage[utf8]{inputenc}
\usepackage[T1]{fontenc}
\usepackage{microtype}

\usepackage{braket}

\usepackage[x11names, rgb, svgnames, table]{xcolor}
\usepackage[inline]{enumitem}

\usepackage[hidelinks]{hyperref}
\usepackage{amsmath}
\usepackage[noabbrev,capitalise]{cleveref}
\usepackage{subcaption}
\usepackage{listings}
\usepackage{graphicx}
\usepackage{tikz}
\usetikzlibrary{shapes.geometric}
\usetikzlibrary{arrows.meta}

\usepackage{biblatex}
\usepackage{csquotes}
\usepackage{breakurl}

\author{Jørgen Kvalsvik \\ \href{mailto:j@lambda.is}{j@lambda.is}}
\title{Modified Condition/Decision Coverage in the GNU Compiler Collection}
\date{January 2, 2025}

\begin{document}
    \maketitle
    \begin{abstract}
  We describe the implementation of the masking Modified
  Condition/Decision Coverage (MC/DC) support in GCC 14, a powerful
  structural coverage metric with wide industry adoption for safety
  critical applications.  By analyzing the structure of Boolean expressions
  with Binary Decision Diagrams we can observe the key property of MC/DC,
  the power to independently affect the outcome, and map to the edges of
  the Control Flow Graph.  This mapping can be translated to a few
  bitwise instructions and enables GCC to instrument programs to
  efficiently observe and record when conditions have been taken and
  have an independent effect on the outcome of a decision.  By analyzing
  the BDD rather than the program syntax, GCC can measure MC/DC for
  almost all of its languages with a single language-agnostic
  implementation, including support for C, C++, D, and Rust.
\end{abstract}

\section{Introduction}
In this paper we describe the algorithms designed and techniques used in
the masking MC/DC support written by the author for the GNU Compiler
Collection (GCC).  Modified Condition/Decision Coverage
(MC/DC)~\cite{hayhurst2001} is a code coverage metric that aims to
improve program quality by requiring that all basic conditions are
shown to have an \emph{independent effect} on the outcome. It has long
been understood that MC/DC is a useful criterion for ensuring high
quality software, notably being mandated by standards such as DO-178
and ISO26262 for software in safety critical systems where malfunction
would put lives at risk.  Development in this space has been slow for
a few decades; the topic of structural coverage and MC/DC received
considerable attention around the
turn of the century~\cite{herring1997,hayhurst2001,chilenski2001,dupuy2000}, but
at the time the analysis was largely manual, a process which is slow
and error prone.  The next decade saw development of automated
instrumentation for analysis~\cite{whalen2013,whalen2015,comar2012}
and the rise of tooling like VectorCAST and LDRA which relies on
source code instrumentation.  GCC has supported statement and branch
coverage since the late 1990s, with GCC the 2.95 manual~\cite{gcc2.95}
describing the gcov tool and coverage. 2019 -- 2024 saw a dramatic shift
in the tooling space, with multiple compiler vendors supporting MC/DC
natively; The Green Hills compiler got native support for MC/DC in
2019~\cite{sagnik2019}, and Clang got support for up-to 6 conditions
MC/DC for C and C++ in 2023~\cite{llvm:D138849}, a restriction which
was relaxed in 2024 \cite{llvm:rfc-76798}.  Support for MC/DC was
introduced in GCC 14, released in 2024.  GCC has always implemented
coverage as object code coverage~\cite{bordin2010,comar2012} (as
opposed to source code coverage), and the MC/DC supports builds on the
object coverage framework by analyzing the Control Flow Graph (CFG).
This is different from the abstract syntax tree driven approaches like
the Whalen et al., Green Hills, and Clang implementations.  There is
also interest in measuring MC/DC of Rust programs~\cite{zaeske2024},
which is supported by the approach described in this paper as the CFG
analysis is unaware of the programming language.

MC/DC is satisfied if:
\begin{itemize}
    \item every entry and exit point has been invoked at least once
    \item every decision in the program has taken all possible outcomes at
        least once
    \item every basic condition has taken on all possible outcomes at least
        once, and
    \item each basic condition has been shown to independently affect the
        decision's outcome
\end{itemize}

There are a few variations on the metric; the most significant ones
are \emph{unique cause} MC/DC and \emph{masking} MC/DC, both described
in detail by Hayhurst et al.~\cite{hayhurst2001}.  Unspecified, it
usually refers to unique cause MC/DC, where only one condition can
vary between two test input vectors to demonstrate independent
decisive power.  Masking MC/DC relaxes this requirement and permits
more than one condition to change between test inputs if they cannot
influence the outcome.  Chilenski~\cite{chilenski2001} demonstrated
they are generally equally effective at finding defects, and
that \emph{masking} MC/DC accepts the largest set of test inputs that
achieves coverage; for $N$ conditions, unique cause MC/DC requires at
least $N+1$ test cases, while masking MC/DC requires $\lceil 2
\sqrt{N} \rceil$ test cases.  In this paper, unless otherwise
specified, MC/DC means \emph{masking} MC/DC.

    \section{Binary Decision Diagrams}
\label{sec:binary-decision-diagrams}
\begin{figure}
    \centering
    \begin{subfigure}[b]{0.25\columnwidth}\centering
  \resizebox{\columnwidth}{!}{\begin{tikzpicture}[>= latex]
  \node [draw, minimum size = 16pt, ellipse]  (a) at (0.5, 2) {$x_1$};
  \node [draw, minimum size = 16pt, ellipse]  (b) at (1, 1)   {$x_2$};
  \node [draw, minimum size = 16pt, rectangle] (c) at (0, 0)   {$1$};
  \node [draw, minimum size = 16pt, rectangle] (d) at (1, 0)   {$0$};

  \path [->]         (a) edge (b);
  \path [->, dotted] (a) edge (c);
  \path [->, dotted] (b) edge (c);
  \path [->]         (b) edge (d);
\end{tikzpicture}}%

        \caption{0 0}
        \label{fig:bdd:or:1}
    \end{subfigure}%
    \begin{subfigure}[b]{0.25\columnwidth}\centering
  \resizebox{\columnwidth}{!}{\begin{tikzpicture}[>= latex]
  \node [draw, minimum size = 16pt, ellipse]  (a) at (0.5, 2) {$x_1$};
  \node [draw, minimum size = 16pt, ellipse]  (b) at (1, 1)   {$x_2$};
  \node [draw, minimum size = 16pt, rectangle] (c) at (0, 0)   {$1$};
  \node [draw, minimum size = 16pt, rectangle] (d) at (1, 0)   {$0$};

  \path [->]         (a) edge (b);
  \path [->, dotted] (a) edge (c);
  \path [->]         (b) edge (c);
  \path [->, dotted] (b) edge (d);
\end{tikzpicture}}%

        \caption{0 1}
        \label{fig:bdd:or:2}
    \end{subfigure}%
    \begin{subfigure}[b]{0.25\columnwidth}\centering
  \resizebox{\columnwidth}{!}{\begin{tikzpicture}[>= latex]
  \node [draw, minimum size = 16pt, ellipse]  (a) at (0.5, 2) {$x_1$};
  \node [draw, minimum size = 16pt, ellipse]  (b) at (1, 1)   {$x_2$};
  \node [draw, minimum size = 16pt, rectangle] (c) at (0, 0)   {$1$};
  \node [draw, minimum size = 16pt, rectangle] (d) at (1, 0)   {$0$};

  \path [->, dotted] (a) edge (b);
  \path [->]         (a) edge (c);
  \path [->, dotted] (b) edge (c);
  \path [->, dotted] (b) edge (d);
\end{tikzpicture}}%

        \caption{1 0}
        \label{fig:bdd:or:3}
    \end{subfigure}%
    \begin{subfigure}[b]{0.25\columnwidth}\centering
  \resizebox{\columnwidth}{!}{\begin{tikzpicture}[>= latex]
  \node [draw, minimum size = 16pt, ellipse]  (a) at (0.5, 2) {$x_1$};
  \node [draw, minimum size = 16pt, ellipse]  (b) at (1, 1)   {$x_2$};
  \node [draw, minimum size = 16pt, rectangle] (c) at (0, 0)   {$1$};
  \node [draw, minimum size = 16pt, rectangle] (d) at (1, 0)   {$0$};

  \path [->, dotted] (a) edge (b);
  \path [->]         (a) edge (c);
  \path [->, dotted] (b) edge (c);
  \path [->, dotted] (b) edge (d);
\end{tikzpicture}}%

        \caption{1 1}
        \label{fig:bdd:or:4}
    \end{subfigure}
    \caption{
        $(x_1 \lor x_2)$ as a BDD. The regular edges make up the path for the
        given input vector. \subref{fig:bdd:or:2} demonstrates masking;
        changing $x_1$ does not change the decision ($1$).
    }
    \label{fig:bdd:or}
\end{figure}
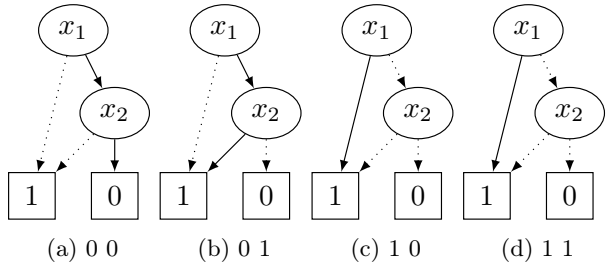

Boolean functions have a canonical representation as a reduced ordered binary
decision diagram (BDD)~\cite{bryant86} where a path from the root to a
\emph{terminal} vertex corresponds to an input vector $(x_1, \ldots, x_n)$, and
the vertices correspond to the evaluation of a basic condition. Reduced,
ordered binary decision diagrams is what is usually meant by BDD, and
all uses of BDD in this paper means reduced ordered binary decision diagram.
Andersen~\cite{andersen99} is a good introduction. A BDD is \emph{ordered} when
all paths through the variables respect the linear order $x_1 < x_2 < \ldots <
x_n$ and \emph{reduced} when there are no redundant tests of variables, and no two
distinct vertices have the same variable name and successors.  Short circuiting
logic has a natural expression in BDDs as shortcut edges to deeper levels,
e.g.\ the edge $(x_1, 1)$ in \cref{fig:bdd:or:1}.  The leaves of a BDD,
called the \emph{terminals} or \emph{decisions}, are denoted with the
literals $0$ and $1$.  The internal vertices, called the
\emph{non-terminals}, correspond to a basic condition in the Boolean
expression and have exactly 2 successors, the then and else
branches. For any BDD there is a path from every $x_i$ to every $x_k$
where $i < k$, and all paths end in one of the two terminals.

Condition coverage is defined as every condition in a Boolean function
(decision) having taken all possible outcomes at least
once~\cite{hayhurst2001}.  Since vertices in the BDD correspond to the
evaluation of the basic conditions, and edges the outcomes of the
basic conditions, we can determine condition coverage by recording the
paths taken through the BDD during execution; coverage is achieved
when every edge has been taken at least once.  In the context of
object code coverage, condition coverage is sometimes called
\emph{edge coverage}.

The key property of MC/DC is the \emph{independence criterion}. For MC/DC,
recording the vertices as they are visited is not sufficient for determining
coverage as some vertices may be visited without independently affecting the
decision due to the \emph{masking effect}.
A condition is \emph{masked} if changing its value while keeping the other
inputs fixed does not change the decision. This is intuitive for short
circuiting - since a short circuited condition is not even evaluated, it
cannot affect the decision. Short circuiting is determined by the left operand
and operator; $(1 \lor x)$ and $(0 \land x)$. Boolean operators are
commutative, $(x_1 \lor x_2) \Leftrightarrow (x_2 \lor x_1)$, so the inability
to affect on the decision must also apply for $(x \lor 1)$ and $(x \land
0)$, i.e.\ the right operand \emph{masks} the left operand. This can be seen in
the truth table in \cref{fig:truth-table:1}; for $(x_1 \lor x_2)$ the
decision can be fully determined by the left operand $x_1 = 1$ (rows 3, 4) and
the $x_2$ does not affect the decision, and likewise for $(x_1 \land
x_2)$ where $x_1 = 0$ (rows 1, 2). For $(x_1 \lor 1)$ (rows 2, 4) and
$(x_1 \land 0)$ (rows 1, 3), the left operand has no effect on the
decision.  \cref{fig:bdd:or} shows the same function as a BDD.
Simply put, conditions are masked if they would be short circuited in
the reverse-order evaluation of the Boolean function; if $(x_i, x)$
short circuits $x_k$, then $(x_k, x)$ masks $x_i$.

\begin{figure}
    \centering
    \begin{subfigure}[b]{0.4\columnwidth}\centering
        \begin{tabular}{c c | c c}
            $x_1$ & $x_2$ & $\lor$ & $\land$ \\
            \hline
            0 & 0 & 0 & 0 \\
            0 & 1 & 1 & 0 \\
            1 & 0 & 1 & 0 \\
            1 & 1 & 1 & 1 \\
        \end{tabular}
        \caption{}
        \label{fig:truth-table:1}
    \end{subfigure}%
    \begin{subfigure}[b]{0.3\columnwidth}\centering
        \begin{tabular}{c c | c}
            $x_1$ & $x_2$ & $\lor$ \\
            \hline
            0 & 0 & 0 \\
            0 & 1 & 1 \\
            1 & * & 1 \\
            1 & * & 1 \\
        \end{tabular}
        \caption{}
        \label{fig:truth-table:2}
    \end{subfigure}%
    \begin{subfigure}[b]{0.3\columnwidth}\centering
        \begin{tabular}{c c | c}
            $x_2$ & $x_1$ & $\lor$ \\
            \hline
            0 & 0 & 0 \\
            1 & * & 1 \\
            0 & 1 & 1 \\
            1 & * & 1 \\
        \end{tabular}
        \caption{}
        \label{fig:truth-table:3}
    \end{subfigure}
    \caption{
        \subref{fig:truth-table:1} is the truth table for the Boolean
        operators.  \subref{fig:truth-table:2} and \subref{fig:truth-table:3}
        are the truth tables for $\lor$ in left-to-right and right-to-left
        evaluation order with conditions short circuited (*).
    }
    \label{fig:truth-table}
\end{figure}

\section{Finding the masking table}
\label{sec:finding-masking-table}
In this section we describe an algorithm for computing the masking table
from the structure of a Boolean function by analyzing the
corresponding BDD.  BDDs were used by Comar et al.~\cite{comar2012} to
show that when the BDD is a tree then edge coverage implies MC/DC,
which GTD GmbH used to implement an MC/DC tool~\cite{gtd:2021} that checks
and requires that all Boolean expressions are tree-like BDDs. This
algorithm only relies on the BDD being reduced and ordered, and works
for arbitrary Boolean functions.

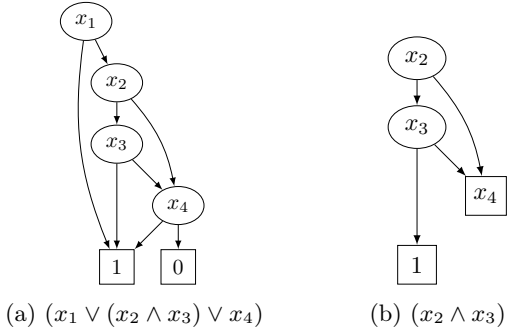
\begin{figure}
    \centering
    \begin{subfigure}[b]{0.5\columnwidth}\centering
  \resizebox{0.5\textwidth}{!}{\begin{tikzpicture}[>= latex]
  \node [draw, minimum size = 16pt, ellipse]  (a) at (0.5, 4) {$x_1$};
  \node [draw, minimum size = 16pt, ellipse]  (b) at (1, 3)   {$x_2$};
  \node [draw, minimum size = 16pt, ellipse]  (c) at (1, 2)   {$x_3$};
  \node [draw, minimum size = 16pt, ellipse]  (d) at (2, 1)   {$x_4$};
  \node [draw, minimum size = 16pt, rectangle] (e) at (1, 0)   {$1$};
  \node [draw, minimum size = 16pt, rectangle] (f) at (2, 0)   {$0$};

  \path [->] (a) edge (b);
  \path [->] (a) edge [bend right = 15] (e);
  \path [->] (b) edge (c);
  \path [->] (b) edge [bend left = 15] (d);
  \path [->] (c) edge (e);
  \path [->] (c) edge (d);
  \path [->] (d) edge (e);
  \path [->] (d) edge (f);
\end{tikzpicture}}%

        \caption{$(x_1 \lor (x_2 \land x_3) \lor x_4)$}
        \label{fig:bdd:pseudo-terminals:1}
    \end{subfigure}%
    \begin{subfigure}[b]{0.5\columnwidth}\centering
  \resizebox{0.5\textwidth}{!}{\begin{tikzpicture}[>= latex]
  \node [draw = none, minimum size = 16pt, ellipse]  (a) at (0.5, 4) {};
  \node [draw, minimum size = 16pt, ellipse]  (b) at (1, 3)   {$x_2$};
  \node [draw, minimum size = 16pt, ellipse]  (c) at (1, 2)   {$x_3$};
  \node [draw, minimum size = 16pt, rectangle]  (d) at (2, 1)   {$x_4$};
  \node [draw, minimum size = 16pt, rectangle] (e) at (1, 0)   {$1$};
  \node [draw = none, minimum size = 16pt, rectangle] (f) at (2, 0)   {};

  \path [->] (b) edge (c);
  \path [->] (b) edge [bend left = 15] (d);
  \path [->] (c) edge (e);
  \path [->] (c) edge (d);
\end{tikzpicture}}%

        \caption{$(x_2 \land x_3)$}
        \label{fig:bdd:pseudo-terminals:2}
    \end{subfigure}

    \caption{
        A Boolean function~(\subref{fig:bdd:pseudo-terminals:1}) and its
        subexpression~(\subref{fig:bdd:pseudo-terminals:2}). The terminals $x_4$
        and $1$ in~(\subref{fig:bdd:pseudo-terminals:2}) become pseudo-terminals
        in~(\subref{fig:bdd:pseudo-terminals:1}). The masking effects in
        subexpressions will be preserved in the superexpression.
    }
    \label{fig:bdd:pseudo-terminals}
\end{figure}

Short circuiting and masking can be understood in the BDD as different
paths to the same terminal. Subexpressions preserve local short
circuiting and masking, seen as multiple paths to a
\emph{pseudo-terminal}, a vertex that would be a terminal if the
expression was independent. Multiple paths to the same pseudo-terminal
means it will have multiple incoming edges (in-degree $\geq 2$); the
short circuiting edges and the one non-short circuiting edge.
Consider the Boolean function $(x_1 \lor (x_2 \land x_3) \lor x_4)$,
and the BDD in \cref{fig:bdd:pseudo-terminals:1}.
$(x_2 \land x_3)$ (\cref{fig:bdd:pseudo-terminals:2}) is
\emph{embedded} in the BDD, and what would be the
terminals in $(x_2 \land x_3)$ become the pseudo-terminals
$x_4$ and $1$, both with an in-degree $\geq 2$. The goal is to
determine which terms are masked upon taking an edge, which
corresponds to a specific outcome of evaluating a condition.  The
specific Boolean operator is generally of little significance for the
analysis, in contrast to the abstract syntax tree approach of
Whalen et al.~\cite{whalen2013}, as the BDDs for the Boolean functions
with inverted operands are isomorphic. This is explained by De
Morgan's laws, $\neg (P \lor Q) \Leftrightarrow (\neg P) \land (\neg
Q)$ and $\neg (P \land Q) \Leftrightarrow (\neg P) \lor (\neg Q)$, and
in the BDD by inverting all the comparisons and swapping the
terminals. Negation is built into the evaluation of a basic condition
(contrast \lstinline{if (v != 0)} to \lstinline{if (v == 0)}).  As a
consequence, we only need to be concerned with the \emph{shape} of the
BDD.

\newcommand{\preds}[1]{\textrm{preds}(#1)}
\newcommand{\succs}[1]{\textrm{succs}(#1)}

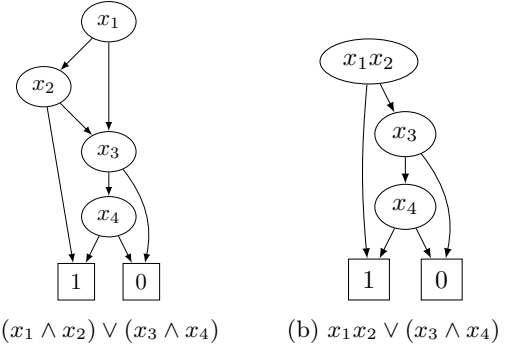
\begin{figure}
    \begin{subfigure}{0.5\columnwidth}\centering
  \resizebox{0.5\textwidth}{!}{\begin{tikzpicture}[>= latex]
  \node [draw, minimum size = 16pt, ellipse]   (a) at (1, 4)   {$x_1$};
  \node [draw, minimum size = 16pt, ellipse]   (b) at (0, 3)   {$x_2$};
  \node [draw, minimum size = 16pt, ellipse]   (c) at (1, 2)   {$x_3$};
  \node [draw, minimum size = 16pt, ellipse]   (d) at (1, 1)   {$x_4$};
  \node [draw, minimum size = 16pt, rectangle] (e) at (0.5, 0) {$1$};
  \node [draw, minimum size = 16pt, rectangle] (f) at (1.5, 0) {$0$};

  \path [->] (a) edge (b);
  \path [->] (a) edge (c);
  \path [->] (b) edge (e);
  \path [->] (b) edge (c);
  \path [->] (c) edge (d);
  \path [->] (c) edge [bend left = 25] (f);
  \path [->] (d) edge (e);
  \path [->] (d) edge (f);
\end{tikzpicture}}%

        \caption{$(x_1 \land x_2) \lor (x_3 \land x_4)$}
    \end{subfigure}%
    \begin{subfigure}{0.5\columnwidth}\centering
  \resizebox{0.5\textwidth}{!}{\begin{tikzpicture}[>= latex]
  \node [draw, minimum size = 16pt, ellipse]   (a) at (0.5, 3)   {$x_1x_2$};
  \node [draw, minimum size = 16pt, ellipse]   (c) at (1, 2)   {$x_3$};
  \node [draw, minimum size = 16pt, ellipse]   (d) at (1, 1)   {$x_4$};
  \node [draw, minimum size = 16pt, rectangle] (e) at (0.5, 0) {$1$};
  \node [draw, minimum size = 16pt, rectangle] (f) at (1.5, 0) {$0$};

  \path [->] (a) edge [bend right = 5] (e);
  \path [->] (a) edge (c);
  \path [->] (c) edge (d);
  \path [->] (c) edge [bend left = 25] (f);
  \path [->] (d) edge (e);
  \path [->] (d) edge (f);
\end{tikzpicture}}%

        \caption{$x_1 x_2 \lor (x_3 \land x_4)$}
    \end{subfigure}
    \caption{
      Two equivalent Boolean functions where a subexpression is
      computed outside the function.  All edges from the subgraph
      $(x_2 \land x_2)$ go to $x_3$ and $1$, which are the successors
      $x_1x_2$.
    }
    \label{fig:complex-complex}
\end{figure}

Consider the Boolean expression $B$ on a normalized form $(A_1 \lor
A_2 \lor \cdots \lor A_n)$ or $(A_1 \land A_2 \land \cdots \land A_n)$
where $A_k$ is either a basic condition or a possibly nested complex
Boolean function combined with a \emph{different} operator.  Let $B_T$
be the short circuit terminal of $B$, $1$ if $\lor$, or $0$ if
$\land$, which will have an in-degree $\geq 2$.  We want to find the
\emph{masking table}, where the notation $(u, v) \sim \set{x_k,
x_{k+1}, \ldots, x_n}$ means that taking the edge $(u, v)$ mask the
conditions $x_k, x_{k+1}, \ldots, x_n$.  $A_{k+1}$ can only be entered
through $A_k$, which follows directly the ordering property.  It
follows that all paths through $A_k$ must go through $A_{k+1}$ or to
$B_T$. For example, let $A_k'$ be the result of evaluating $A_k$ on
its own.  Precomputing would not change the truth table of $B$, and
since $A_k'$ is a basic condition it has exactly two successors, one
being the short circuiting edge to $B_T$, and the other the evaluation
of the right operand $A_{k+1}$. An example can be seen in
\cref{fig:complex-complex} where the substitution $x_1x_2 = (x_1
\land x_2)$ is succeeded by $1$ and $x_3$, which corresponds to the
edges $(x_1, x_3), (x_2, x_3), (x_2, 1)$.  Finally, the last condition
of $A_k$ \emph{must} decide the outcome.  By using these observations
we can identify the vertices of $A_k$ from $B$; given the edges $(A_i,
B_T)$ and $(A_k, B_T)$ where $i < k$, $(A_k, B_T)$ masks $A_i$.  The
problem now becomes finding the subset of vertices where all paths go
through either \begin{enumerate*}[
  label = (\arabic*),
  itemjoin = {{, }},
  itemjoin* = {{, or }}]
  \item the edge $(A_i, B_T)$, where the source vertex is last term of
    $A_i$
  \item an edge to the term of $A_{i+1}$
  \end{enumerate*}.  The boundaries of $A_k$ can be found with the function
$P(x) = \{(x_e, x_n, x_m) \mid
\allowbreak x_n \in \preds{x},
\allowbreak x_m \in \preds{x},
\allowbreak x_e \in \succs{x} - \set{x_n},
\allowbreak n < m\}$; for each vertex $x$ with an in-degree $\geq 2$,
consider the Cartesian product of the predecessors $\preds{x}^2$ where
the pairs $(x_n, x_m)$ are ordered so that $n < m$ and duplicates are
removed, and $x_e$ is the non-$x$ successor of $x_n$.  Note that
$\succs{A_k} = \set{x, x_e}$.  Masked conditions can found from the
triple $(x_e, x_n, x_m)$: if all paths from a vertex $x_i$ where $i <=
n$ go through either $x_e$ or $x$ it is masked when the edge $(x_m,
x)$ is taken.

\begin{figure}
    \centering
    \begin{subfigure}[b]{0.50\columnwidth}\centering
  \resizebox{0.4\textwidth}{!}{\begin{tikzpicture}[>= latex]
  \node [draw, minimum size = 16pt, ellipse]  (a) at (0, 3) {$x_1$};
  \node [draw, minimum size = 16pt, ellipse]  (b) at (0.5, 2) {$x_2$};
  \node [draw, minimum size = 16pt, ellipse]  (c) at (1, 1) {$x_3$};
  \node [draw, minimum size = 16pt, rectangle] (d) at (0, 0)   {$1$};
  \node [draw, minimum size = 16pt, rectangle] (e) at (1, 0)   {$0$};

  \path [->] (a) edge (d);
  \path [->] (a) edge (b);
  \path [->] (b) edge (d);
  \path [->] (b) edge (c);
  \path [->] (c) edge (d);
  \path [->] (c) edge (e);
\end{tikzpicture}}%

        \caption{}
    \end{subfigure}%
    \begin{subfigure}[b]{0.5\columnwidth}\centering
        \begin{tabular}{c | c c c}
            & $x_1$ & $x_2$ & $x_3$ \\
            \hline
            $x_1 = 1$ &   & * & * \\
            $x_2 = 1$ & - &   & * \\
            $x_3 = 1$ & - & - &   \\
        \end{tabular}
        \caption{}
        \label{fig:bdd:or-3:table}
    \end{subfigure}
    \caption{
        $(x_1 \lor x_2 \lor x_3)$ and the short circuited (*) and masked (-)
        terms when a basic condition takes on 1.
    }
    \label{fig:bdd:or-3}
\end{figure}
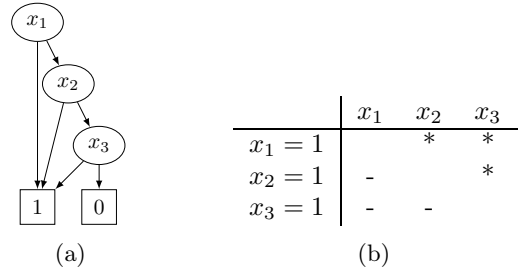

We now have a an algorithm for computing the masking table from the
BDD:
\begin{enumerate}
    \item For each vertex $x$ with in-degree $\geq 2$, find the triples
      $\set{(x_e, x_n, x_m), \ldots} = P(x)$.
    \item For each triple, remove $\succs{x_n} = \set{x, x_e}$ from
      the BDD; these are the pseudo-terminals of $A_k$.
    \item Remove and collect all vertices made into leaves until no
      more vertices can be removed; the collected vertices are $A_k$.
    \item Add the collected vertices to the masking table; $M(x_m, x)
      + A_k$.
\end{enumerate}

An example run of the algorithm can be seen in
\cref{fig:steps-masked-conditions}.  $A_k$ may be an arbitrarily
complex BDD, but since all paths must go through either $x$ or $x_e$
and removing leaves is equivalent to inverting the edges and collecting
all paths from $x$ and $x_e$ to the first condition in $A_k$,
e.g.\ with a breadth-first search.  The search will stop when it
reaches the root of the BDD or the vertex $x_p$ where there is a path
from $x_p$ to the other pseudo-terminal of $A_k$.  Note that algorithm
may collect multiple $A_i$ where $i < k$ that short circuit to the
same pseudo-terminal $x$.  The simplest example is $(x_1 \lor x_2 \lor
x_3)$, as seen in \cref{fig:bdd:or-3}.  For this function,
$P(1)$ contains $(x_e, x_n, x_m) = (x_3, x_2, x_3)$, which means $1$
and $x_3$ should be removed from the graph.  This makes $x_2$ a leaf,
which when removed makes $x_1$ a leaf and yields the masking table
entry $(x_3, 1) \sim \set{x_1, x_2}$.  The correctness is not affected
as the masking table maps edges to sets with duplicates removed, and
the inefficiency can be addressed with memoization.  It is possible
that there is an even faster approach for this, either by limiting the
search by going through the other pseudo-terminal, or by removing all
$x_e$ at once, but these approaches were not explored.

\begin{figure}
    \begin{subfigure}{0.25\columnwidth}\centering
  \resizebox{\columnwidth}{!}{\begin{tikzpicture}[>= latex]
  \node [draw, minimum size = 16pt, ellipse]   (a) at (1, 5)   {$x_1$};
  \node [draw, minimum size = 16pt, ellipse]   (b) at (1.5, 4) {$x_2$};
  \node [draw, minimum size = 16pt, ellipse]   (c) at (0.5, 3) {$x_3$};
  \node [draw, minimum size = 16pt, ellipse]   (d) at (0.5, 2) {$x_4$};
  \node [draw, minimum size = 16pt, ellipse]   (e) at (0, 1)   {$x_5$};
  \node [draw, minimum size = 16pt, rectangle] (f) at (0, 0)   {$1$};
  \node [draw, minimum size = 16pt, rectangle] (g) at (1, 0)   {$0$};

  \path [->] (a) edge (b);
  \path [->] (a) edge (c);
  \path [->] (b) edge (c);
  \path [->] (b) edge [bend left = 5] (g);
  \path [->] (c) edge (d);
  \path [->] (c) edge [bend right = 25] (e);
  \path [->] (d) edge (e);
  \path [->] (d) edge (g);
  \path [->] (e) edge (f);
  \path [->] (e) edge (g);
\end{tikzpicture}}%

        \caption{}
        \label{fig:steps:1}
    \end{subfigure}%
    \begin{subfigure}{0.25\columnwidth}\centering
  \resizebox{\columnwidth}{!}{\begin{tikzpicture}[>= latex]
  \node [draw, minimum size = 16pt, ellipse]           (a) at (1, 5)   {$x_1$};
  \node [draw, minimum size = 16pt, ellipse]           (b) at (1.5, 4) {$x_2$};
  \node [draw, minimum size = 16pt, ellipse, dotted]   (c) at (0.5, 3) {$x_3$};
  \node [draw, minimum size = 16pt, ellipse]           (d) at (0.5, 2) {$x_4$};
  \node [draw, minimum size = 16pt, ellipse]           (e) at (0, 1)   {$x_5$};
  \node [draw, minimum size = 16pt, rectangle]         (f) at (0, 0)   {$1$};
  \node [draw, minimum size = 16pt, rectangle, dotted] (g) at (1, 0)   {$0$};

  \path [->] (a) edge (b);
  \path [->, dotted] (a) edge (c);
  \path [->, dotted] (b) edge (c);
  \path [->, dotted] (b) edge [bend left = 5] (g);
  \path [->] (c) edge (d);
  \path [->] (c) edge [bend right = 25] (e);
  \path [->] (d) edge (e);
  \path [->, dotted] (d) edge (g);
  \path [->] (e) edge (f);
  \path [->, dotted] (e) edge (g);
\end{tikzpicture}}%

        \caption{}
        \label{fig:steps:2}
    \end{subfigure}%
    \begin{subfigure}{0.25\columnwidth}\centering
  \resizebox{\columnwidth}{!}{\begin{tikzpicture}[>= latex]
  \node [draw, minimum size = 16pt, ellipse, dotted]   (a) at (1, 5)   {$x_1$};
  \node [draw, minimum size = 16pt, ellipse, dotted]   (b) at (1.5, 4) {$x_2$};
  \node [draw, minimum size = 16pt, ellipse, dotted]   (c) at (0.5, 3) {$x_3$};
  \node [draw, minimum size = 16pt, ellipse]           (d) at (0.5, 2) {$x_4$};
  \node [draw, minimum size = 16pt, ellipse]           (e) at (0, 1)   {$x_5$};
  \node [draw, minimum size = 16pt, rectangle]         (f) at (0, 0)   {$1$};
  \node [draw, minimum size = 16pt, rectangle, dotted] (g) at (1, 0)   {$0$};

  \path [->, dotted] (a) edge (b);
  \path [->, dotted] (a) edge (c);
  \path [->, dotted] (b) edge (c);
  \path [->, dotted] (b) edge [bend left = 5] (g);
  \path [->] (c) edge (d);
  \path [->] (c) edge [bend right = 25] (e);
  \path [->] (d) edge (e);
  \path [->, dotted] (d) edge (g);
  \path [->] (e) edge (f);
  \path [->, dotted] (e) edge (g);
\end{tikzpicture}}%

        \caption{}
        \label{fig:steps:3}
    \end{subfigure}%
    \begin{subfigure}{0.25\columnwidth}\centering
  \resizebox{\columnwidth}{!}{\begin{tikzpicture}[>= latex]
  \node [draw, minimum size = 16pt, ellipse, dotted]   (a) at (1, 5)   {$x_1$};
  \node [draw, minimum size = 16pt, ellipse, dotted]   (b) at (1.5, 4) {$x_2$};
  \node [draw, minimum size = 16pt, ellipse]           (c) at (0.5, 3) {$x_3$};
  \node [draw, minimum size = 16pt, ellipse]           (d) at (0.5, 2) {$x_4$};
  \node [draw, minimum size = 16pt, ellipse]           (e) at (0, 1)   {$x_5$};
  \node [draw, minimum size = 16pt, rectangle]         (f) at (0, 0)   {$1$};
  \node [draw, minimum size = 16pt, rectangle]         (g) at (1, 0)   {$0$};

  \path [->, dotted] (a) edge (b);
  \path [->, dotted] (a) edge (c);
  \path [->, dotted] (b) edge (c);
  \path [->, dotted] (b) edge [bend left = 5] (g);
  \path [->] (c) edge (d);
  \path [->] (c) edge [bend right = 25] (e);
  \path [->] (d) edge (e);
  \path [->] (d) edge (g);
  \path [->] (e) edge (f);
  \path [->] (e) edge (g);
\end{tikzpicture}}%

        \caption{}
        \label{fig:steps:4}
    \end{subfigure}
    \caption{
      $((x_1 \lor x_2) \land (x_3 \lor x_4) \land x_5)$ for the
      edge $(x_4, 0)$ and $(x_e, x_n, x_m) = (x_3, x_2, x_4)$.
      In the first step (\subref{fig:steps:2}) $\succs{x_n} =
      \set{x_3, 0}$ are removed (dotted).  Repeatedly removing leaves
      gives (\subref{fig:steps:3}). (\subref{fig:steps:4}) shows the
      collected conditions; $(x_4, 0) \sim \set{x_1, x_2}$.
    }
    \label{fig:steps-masked-conditions}
\end{figure}

\section{Instrumenting programs for measuring MC/DC}
If the control flow graph (CFG) is carefully constructed to directly
model the evaluation of conditional expressions as BDDs this algorithm
can be run directly on the CFG.  This is a departure from the
approaches of Whalen et al.~\cite{whalen2013} and
Sagnik~\cite{sagnik2019} who derive the masking table from the
abstract syntax tree (AST).  By doing the analysis directly on the CFG
the analysis becomes language agnostic, or requires minimal
information from the front-end, and can be used with several languages.
For this work the compilers for C, C++, D, and Rust (which is
experimental in GCC 14) were all shown capable to instrument for
MC/DC.  Note that the Go front-end does not construct a CFG isomorphic
to the canonical BDD for Boolean expressions and consequently cannot
measure MC/DC.  Other languages supported by GCC (Ada, Fortran,
Objective-C, Modula-2) were not tested, but might work.
The only modification other than the CFG analysis was
in a lowering pass, where complex Boolean expressions were transformed
to If-then-else normal form (INF), which was extended with an
identifier that maps a basic condition to its Boolean expression.  INF
is a Boolean expression built entirely from the if-then-else operator
and the constants $0$ and $1$ such that all tests are performed on
variables.  Any Boolean function is expressible in
INF~\cite{andersen99} which can be represented in graph form as a
\emph{decision tree} and in a refined form a BDD.  The BDD is a
natural way for compilers to implement Boolean expression evaluation
as it encodes both evaluation order and short circuiting with no
redundant tests.

The approach is similar to measuring condition coverage as described
in \cref{sec:binary-decision-diagrams} by recording the paths
through the BDD during execution. However, in MC/DC an edge may be
taken without having an independent effect on the
decision for that input vector. The masking table is a function $m : E
\to 2^{\mathcal{C}}$ where $\mathcal{C}$ is the set of basic
conditions.  Then $m : e \mapsto C \subseteq \mathcal{C}$ maps $e$ to
a possibly empty set of basic conditions that do not have an effect
on the decision.
When the program reaches a terminal through a path $E$, the covered
condition outcomes will be the edges $E - \set{x : m(E)}$, that is,
with the contribution of the masked conditions voided and the
remaining condition outcomes (edges) shown to have an independent
effect.  Coverage is achieved when all edges have been marked at least
once. For example, given $((x_1 \lor x_2) \land (x_3 \lor x_4) \land
x_5)$ in \cref{fig:steps-masked-conditions}, the masking table
$m$ in \cref{fig:masking-table}, and the input vector
$(0~1~0~0~1)$. The edges taken are
$(x_1,x_2)~(x_2,x_3)~(x_3,x_4)~(x_4,0)$. Applying $m$ to the edges
gives the $\set{x_1, x_2}$ which means the covered condition outcomes
are $x_3 = 0, x_4 = 0$. For condition coverage the covered outcomes
would be $x_1 = 0,~x_2 = 1,~x_3 = 0,~x_4 = 0$.

\begin{figure}
    \begin{tabular}{l l l}
        Edge & Masked conditions            & Bitmask \\
        $(x_2, x_3)$ & $x_1$                & 10000 \\
        $(x_4, x_5)$ & $x_3$                & 00100 \\
        $(x_4, 0)$   & $x_1, x_2$           & 11000 \\
        $(x_5, 0)$   & $x_1, x_2, x_3, x_4$ & 11110 \\
    \end{tabular}
    \caption{
        Masking table $m$ for $((x_1 \lor x_2) \land (x_3 \lor x_4) \land x_5)$
        (\cref{fig:steps-masked-conditions}). The third column is the
        bitmask representation of the masked conditions.
    }
    \label{fig:masking-table}
\end{figure}

As seen in \cref{sec:finding-masking-table} the masking table
$m$ only depends on the structure of the Boolean function,
and can be computed offline, i.e.\ during compilation.  There is a bijection
between the position (index) of the basic conditions in a Boolean function and
the natural numbers. By limiting the number of conditions, paths can be
represented and recorded using fixed-size bitsets and
bitwise operations. For example, the input vector $(0~1~0~0~1)$
and path $(x_1,x_2)~(x_2,x_3)~(x_3,x_4)~(x_4,0)$ can be represented as the two
bitsets $f = 10110, t = 01000$ where bit $t[n] = 1$ if the $n$th condition was
true, $f[n] = 1$ for false. Note that $t[5] = f[5] = 0$ because $x_5$ was short
circuited.  Edges are recorded and masked by with a few bitwise
instructions before performing the conditional jump, as seen in
\cref{fig:instr-cond-jmp}.  Upon taking an edge to a terminal
both bitsets are flushed to global bitsets, what Whalen et
al.~\cite{whalen2013} call \emph{independence arrays}. Coverage is
achieved when all the bits are set in both
counters. Sagnik~\cite{sagnik2019} uses a similar approach with a
single bitset, using odd/even indices for the true/false outcomes.
Limiting the number of conditions and using fixed-size bitsets is a
purely practical choice, and the approach would work just as well with
variable sized bitsets. In GCC 14, the bitset is 32 or 64 depending on
the target platform and configuration. Boolean functions with more
than 32 conditions are very uncommon, 64 even more so, so this
limitation rarely becomes an issue.

\begin{figure}
    \begin{lstlisting}[basicstyle = \ttfamily]
if (x_n != 0)
  then: goto _then_n
  else: goto _else_n

if (x_n != 0)
  then:
    _t &= ~m[n, 1]
    _f &= ~m[n, 1]
    _t |= (1 << n)
    goto _then_n
  else:
    _t &= ~m[n, 0]
    _f &= ~m[n, 0]
    _f |= (1 << n)
    goto _else_n
    \end{lstlisting}
    \caption{
        The evaluation of a basic condition in INF with and without
        instrumentation.  \lstinline{m[n, 0]} looks up the bitmask for the
        false outcome of the $n$th condition in $m$. Operators use the
        semantics from C: \lstinline{&=} is bitwise-and,
        \lstinline{|=} is bitwise-or, \texttt{\textasciitilde} inverts
        all bits, and \lstinline{1 << n} is a bitmask where only the
        $n$th bit is set.
    }
    \label{fig:instr-cond-jmp}
\end{figure}

\section{Summary and future work}
Interpreting the Control Flow Graph (CFG) directly as a Binary
Decision Diagram (BDD) is an effective way of inferring the masking-
and independence properties of Boolean expressions.  The novel
implementation of Modified Condition/Decision Coverage (MC/DC) in the
GNU Compiler Collection (GCC) does not build on syntax, but rather
analyzes the BDD and builds a table that maps decisions to efficient
operations on bitsets.  These operations are inserted into the
instrumented program and records when a condition is evaluated and
shown to have an independent effect on the outcome, per the masking
MC/DC criterion.  The BDD analysis is language independent unlike the
syntax based approach taken by Green Hills~\cite{sagnik2019},
Clang~\cite{llvm:D138849}, and Whalen et al.~\cite{whalen2013}, and
the implementation can be shared between C, C++, D, Rust, and any
language where the compiler front-end generates a BDD-like CFG for
Boolean expressions.

The GCC implementation is limited by the size of the bitset, typically
64 bits (one bit per basic condition). This constraint could be
relaxed, either by tuning the algorithm or by targeting variable-sized
bitsets.  The approach also relies on the compiler front-end encoding
Boolean expressions as BDDs, which the Go front-end notably does not,
and future work may either detect whether the CFG is a BDD.  Future
work may also extend GCC to support the other forms of MC/DC, in
particular unique cause MC/DC.

    \printbibliography
\end{document}